\documentclass[prb,aps,twocolumn,groupedaddress,floatfix,citeautoscript,showpacs]{revtex4-2}
\usepackage{graphicx,rotating,subfigure,amsmath,amsfonts,amssymb,delarray,color}
\usepackage{dsfont}
\usepackage{multirow}
\usepackage{comment}

\def\12{\frac{1}{2}}

\graphicspath{{figs/}}

\begin{document}
\bibliographystyle{apsrev}
\title{Entropy dynamics of the binary bond disordered Heisenberg chain}
\author{Di Han}
\affiliation{Hebei Key Laboratory of Photophysics Research and Application, Hebei Normal University, Shijiazhuang, Hebei 050024, China}
\author{Yankui Bai}
\email{ykbai@hebtu.edu.cn}
\affiliation{Hebei Key Laboratory of Photophysics Research and Application, Hebei Normal University, Shijiazhuang, Hebei 050024, China}
\author{Yang Zhao}
\email{zhaoyang22@hebtu.edu.cn}
\affiliation{Hebei Key Laboratory of Photophysics Research and Application, Hebei Normal University, Shijiazhuang, Hebei 050024, China}

\date{\today}
\begin{abstract}
In this article, we study the quench dynamics of the binary bond disordered Heisenberg spin chain. First, we develop a new algorithm, the ancilla TEBD method, which combines the purification technique and the time-evolving block decimation (TEBD) algorithm to study the entanglement dynamics of binary bonded disordered spin chains. With the support of exact diagonalization (ED), we calculate the multifaractal dimension of the binary bond disordered Heisenberg spin model and study its dependence on the strength of the disorder potential; we find that the multifaractal dimension shows no critical behavior which rules out the existence of the many body localization transition. Then, we reproduce the long time scaling of the von Neumann entropy at the time scale that is beyond the reach of typical TEBD and time dependent density matrix renormalization group (tDMRG) algorithms. Based on the numerical analysis, we propose that such a long time scaling is due to the competition of the spin interaction and the disorder which can be seen as a new mechanism for the generating of long time scale entropy dynamics. At last, we numerically proved the existence of the transient Mpemba effect in the bond disordered Heisenberg chain. 
\end{abstract}


\maketitle

\section{Introduction}\label{intro}
The external disorder potential, especially for the onsite disorder, can localize, at least partially, the eigenstates of a closed quantum many body system. For the non-interacting many particle systems, all the eigenstates of low dimensional models (with space dimension $d\leq2$) can be Anderson localized within a finite onsite disorder potential; while, for $d>2$, mobility edges splitting the localized and delocalized states in the spectrum emerge. Even with the existence of many body interactions, eigenstates of a many body system at the edge of the energy spectrum can still be putatively localized by a strong enough disorder potential; the localized states are called many body localized (MBL) states. 

A many body localized quantum state obeys entanglement area law and bears extensive numbers of local integral of motion (LIOM). Usually, the ground state of a quantum many body system with a finite gap to its first excitation state yields the entanglement area law; however, in a fully many body localised system, every eigenstate obeys the area law which gives rise to the long-time logarithmic scaling of the entanglement entropy and the subdiffusive relaxation of the particle number distribution or imbalance\cite{PImb}. For a partially many body localised system, as in the Anderson localized model when $d=3$, there also exists mobility edges in the energy spectrum. The dynamical properties of such partially MBL systems rely on the characteristics of both of the localized and delocalized eigenstates; and the situation becomes more complex when non-abelian symmetry is imposed on the disordered system.

It may be generally accepted that the continuous non-abelian symmetry is not compatible with the many body localization\cite{0su2,1stsu2,Abaninprx,andreas2024}. Non-abelian symmetry, e.g. $SU(2)$, usually results in extensive degeneracy in the energy level spectrum even at the present of a strong random-bond disorder; eigenstates with similar energy density can produce strong resonances that can drive the whole system to thermalization eventually. Even though the many body localized state can not survive in a non-abelian symmetric model, some of the excited eigenstates in such a system can be still subthermal, e.g. a quantum critical glass state (QCG)\cite{qcg,Abaninprx}; the excited states feature the faster entanglement scaling law: $S(L)\sim c\log(L)$\cite{Abaninprx}, comparing to the area law. Also, one can not rule out the existence of some kind of non-thermal and meanwhile non-MBL states; non-trivial entanglement dynamics may as well be introduced by such novel ergodicity breaking states and need to be confirmed by numerical calculations. 

The prototype disordered model with non-abelian symmetry is the bond disordered Heisenberg chain. Previously, the dynamical properties of the bond disordered XXZ spin model has been studied by utilizing many methods such as the generalized real space renormalization group for dynamical systems (RSRG-X)\cite{Abaninprx,XXZMBL2024}, quantum Monte Carlo (QMC)\cite{Vojtaprb}, where the non-abelian symmetry is absent; all of these works predicate a logarithmic like entanglement entropy growth $S(t)\sim\ln^\alpha(t)$, where $\alpha$ is non-universal. It is challenging to explicitly fix the value of $\alpha$ as well as its tight relations with the model's space dimension and disorder strength. Then, for the $SU(2)$ symmetric bond disordered Heisenberg chain, the dynamical properties has to be obtained numerically since the strong resonances between degenerated eigenstates render the RSRG predictions unavailable. The time dependent density matrix renormalization group (tDMRG) algorithm shows that $S(t)\sim\ln(t)$ when $t<10^2$; however, ED results present another logarithmic scaling behavior\cite{zy16} at a much longer time scale $10^2<t<10^3$ which is beyond the validity of tDMRG method due to the proliferation of Trotter error; and this special long time scaling has not been well explained yet. Also, this unusual long time scaling behaviors indicate that the relation between the real time dynamics and the non-ergodic eigenstates of the $SU(2)$ disordered Heisenberg model need to be further explored as well.

In this article, by combining the TEBD algorithm and purification scheme, we developed an efficient time evolving method in order to do the summation with respect to a large number of disorder samples for the real time dynamical properties of binary bond disordered Heisenberg model. Numerically, we generalize the light cone renormalization group (LCRG) algorithm and the purification method\cite{felix} which is initially designed to simulate the potential disorder; We combine the same purification method with the time evolving block decimation (TEBD) algorithm to study the bond disordered spin chain so that we can consider all the binary bond disorder samples in only one cycle of time evolution process. We will use this new method, which is dubbed as the ancilla TEBD, to study the quench dynamics of the von Neumann entropy $S(t)$ of the bond disordered Heisenberg spin chain; and we find that this new algorithm works efficiently up to time scale $t\sim 10^2$. However, to use the ancilla TEBD method, we have to double the chain length which inevitably increases the computing cost. In order to have a check of the long time scaling of $S(t)$ for a longer chain length $L$ and speed up the calculation, we can focus on the dynamical properties in the subspace with non zero total spin moment $S_z^{tot}=-L/2+2$ where the interacting effects are encoded in the up-spin pair. 

Combining with the ED method, we give a through study of the entropy dynamics of the binary bond disordered Heisenberg model. We reproduced the long time scaling behavior of the von Neumann entropy $S(t)$ and ascertain that the special scaling is associated to the distribution of the weak and strong disordered bonds in the potential; it is not related to the localization of eigenstates. Our conclusions are in line with previous tDMRG and ED results. Additionally, using the ancilla TEBD algorithm, we also calculate the entanglement asymmetry, $\Delta S_A(t)$\cite{Mpemba1} in the bond disordered Heisenberg model. It has been proposed that there exists strong memory effect\cite{Mpembag} in the states of spin glass systems which is in tight relation with disordered spin models; and such memory effect can induce the famous Mpemba effect which tells one initial state that is farther from the equilibrate state will relax more faster than other states that is relatively nearer to the equilibrium. Interestingly, in the short time domain, we find that the relaxing rate of $\Delta S_A(t)$ of the two initially asymmetric state is not monotonic and intersects during the time evolution which mimics the quantum Mpemba effect. 

To give a deep insight of the entropy dynamics of the bond disordered Heisenberg chain, we also investigate the mutifractality of its spectrum. Since the eigenstates of the bond disordered Heisenberg model are neither MBL nor purely ergodic, they might be classified as the nonergodic extended (NEE) states\cite{MuF3} whose $q$ moments of the wavefunction amplitudes scales with the chain length $L$ as $\sum_{j=1}^{L}|\phi(j)|^{2q}\sim L^{-D_q(q-1)}$\cite{MuF2}, $D_q$ is called the multifractal dimension; $D_q=1$ ($=0$) means the state is fully extended (localized). A multifractal eigenstate usually occupies a fractal part of the whole Hilbert space; therefore, it is non-ergodic but still extended. The existence of such multifractal states can moderately affected the system's physical properties\cite{MuF4,MuF1,MuF2,MuF3}, such as the ballistic spreading of the initially localized wave packet and the subdiffusive transportation\cite{Geisel97}. For the bond disordered Heisenberg model, its multifractal dimension can be extracted from the scaling behavior of the paricipation entropy $S_q$ (logarithm of the inverse participation ratio (IPR)) with the size of the Hilbert space: $S_q\sim D_q\ln D_H$\cite{laflo19}. We will show that $D_q$ of the bond disordered Heisenberg model is very near to $1$ and independent to the disorder strength; in contrast, $D_q$ exhibit a step jump with respect to the disorder strength for the potential disordered model.

Further, to get a dynamical view of the time evolved initial product wave function $|\Phi(t=0)\rangle$, we also compute the time dependent fidelity $F(t)=|\langle\Phi(0)|\Phi(t)\rangle|^2$ and the instant IPR $I_{IPR}(t)$ of $|\Phi(t)\rangle$. We find that $F(t)$ exhibit robust oscillations in the time domain where the entanglement entropy shows logarithmic scaling; we may infer that the onset of logarithmic scaling is also related to the action of LIMOs. As for $I_{IPR}(t)$ in the spin configuration space, it always converges to a non-universal finite number which indicates the extent of localization of the long time final state; Based on the ED results, we argue that the final state of the $SU(2)$ symmetric disordered Heisenberg model at the long time limit is neither localized nor fully extended no matter how strong the disorder strength is. Our results proves that the long time scaling behavior of $S(t)$ can also be achieved in a non-MBL spin chain.

This article is organized as follows: In Sec.\ref{s_model}, we define the model and introduce the ancilla TEBD algorithm. Then, we compare the ancilla TEBD and ED results to prove the correctness of our new algorithm. In Sec.\ref{multi}, we will calculate the multifractal dimension $D_q$ of the bond disordered Heisenberg model and show the dependence of $D_q$ on the disorder strength. Next, we study the unusual long time scaling of $S(t)$ using ED and the ancilla TEBD method and try to address its physical origin numerically. In Sec.\ref{Mpemba}, we report the decaying behavior of the entanglement asymmetry. Finally, in Sec.\ref{conclude}, we conclude the paper. 

\section{Hamiltonian and method}\label{s_model}
Generally, a bond disordered spin chain is defined as
\begin{eqnarray}\label{eq0}
   H&=& \sum_iD_i[J(S^{+}_iS^{-}_{i+1} + h.c.)+\Delta S^z_iS^z_{i+1}]\nonumber,
\end{eqnarray}
here $D_i$ is uniformly distributed in $[-W/2,W/2]$, $W$ is the disorder strength. For the binary disorder potential, $D_i$ picks each of the values $\pm W/2$ with equal probability; and the corresponding Hamiltonian reads
\begin{eqnarray}\label{eq1}
   H&=& \sum_i[J(1\pm (-)^iD)(S^{+}_iS^{-}_{i+1} + h.c.)\\
   &+&\Delta(1\pm (-)^iD) S^z_iS^z_{i+1}]\nonumber.
\end{eqnarray}
Then, following the procedure of Ref.[\onlinecite{felix}], Eq.\ref{eq1} can be rewritten as the dynamically equivalent formalism  
\begin{eqnarray}\label{eq2}
   H&=& \sum_i[J(S^{+}_iS^{-}_{i+1} + h.c.)+\Delta S^z_iS^z_{i+1}]\nonumber\\
   &+&\sum_i\frac{W}{2}[J(\sigma^z_iS^{+}_iS^{-}_{i+1} + h.c.)+\Delta \sigma^z_iS^z_iS^z_{i+1}].
\end{eqnarray}
Here $\sigma^z$ is the Pauli matrix representing the ancilla site. We choose the initial state as $|\psi_0\rangle=\prod_i|\uparrow or\downarrow\rangle_i(|\uparrow \rangle_i+|\downarrow\rangle_i)_{anci}/\sqrt{2}$. Writting the initial wave-function in this way, we can see that the expectation of the ancilla operator $\langle\sigma^z_i\rangle$ acquires the value of $\pm1$ with equal probability which means that the local physical magnetic moment $S^z$ is imposed upon a binary random field which picks values randomly from $[-W/2,W/2]$ during the time evolution process. In this way, we can simulate the total $2^{L-1}$ disorder samples in only one time evolving process; This technique has been successfully implemented in Ref.[\onlinecite{felix}] for the case of potentially disordered spin chains. Here, we will concentrate instead on the $1$D Heisenberg model with random bonds.

\begin{figure}
  \centering
  \includegraphics[width=1.0\linewidth]{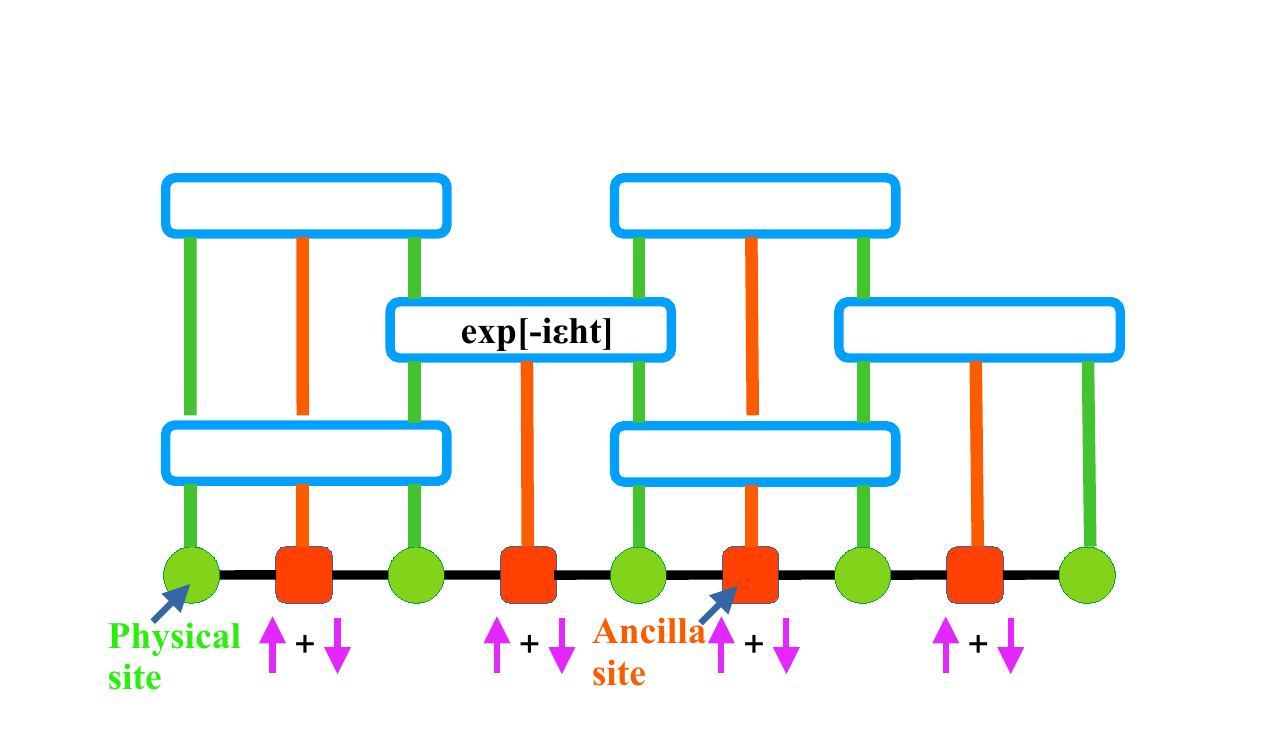}
  \caption{The scheme of Trotter-Suzuki decomposition for the bond disordered Heisenberg chain with ancilla sites. The physical spin sites (physical sites) are plotted as green balls and the ancilla sites as red squares.}\label{fig0}
\end{figure}

To implement the purification trick in TEBD algorithm, we first decompose the time evolution operator $\exp(-iHt)$, here $H$ is represented as Eq.\ref{eq2}, into a Trotter network as shown in Fig.\ref{fig0}. To keep the symmetrical form of the matrix product state (MPS), we arrange the physical spins and ancilla sites alternatively along the chain. In this way, the local time evolution operators $\exp(-i\varepsilon h_{m,\sigma,n}t)$, $\varepsilon$ is the Trotter slice, include three spin half degrees: $\hat{S}_m$, $\hat{S}_n$ and $\sigma$; the local Hilbert space is enlarged by a factor of $2$ comparing to the typical TEBD method. Also, the length of the MPS is also prolonged by the addition of ancilla sites. In this paper, to balance the system size and the temporal cost of calculating, we use the first order Trotter-Suzuki decomposition with $\varepsilon=0.1$ and choose chain length $L$ from $20$ to $60$ according to different scenarios. The truncating dimension $D_c$ is chosen from $100$ to $400$ in accordance with the chosen $L$ and the time scale of the evolution. Next, we will first compare our numerical simulations using short spin chains with ED in the following section to confirm the correctness of our new algorithm.

\begin{figure}
  \centering
  \includegraphics[width=1.0\linewidth]{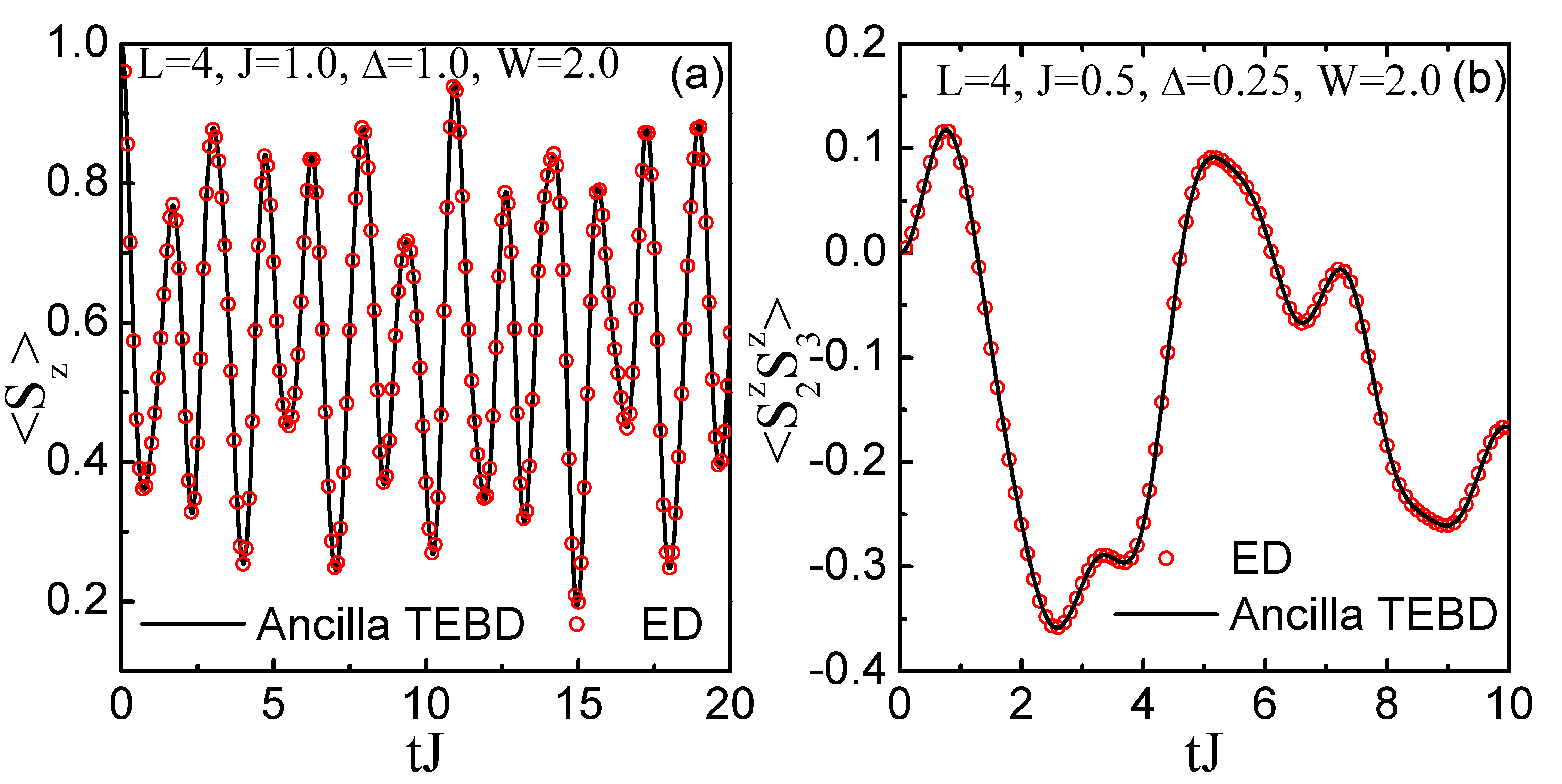}
  \caption{Comparison with the ED method; (a) shows the time evolving of the local spin moments in the middle of a bond disordered XXZ chain with length $L=4$, other parameters are shown in the panel; (b) plots the time dependent spin correlation of the bond disordered Heisenberg model.}\label{fig1}
\end{figure}

Fig.\ref{fig1} plots the comparison of the results of ED and ancilla TEBD. We choose the bond disordered spin chain of length $L=4$ as the testing model whose full disorder sample number is $2^{L-1}=8$. So we can have a check of the validity of the ancilla TEBD by comparing with ED results that utilize full number disorder sample average. For a 1D XXZ chain with $L=4$ and a Neel initial state, see Fig.\ref{fig1} (a), the local spin moment $S_{L/2}^z$ obtained from the two methods are exactly coincide with each other. Fig.\ref{fig1} (b) shows the value of $\langle S^z_2S^z_{3}\rangle$ between two singlets when a product singlet initial state $(1/2)(|\uparrow\downarrow\rangle+|\downarrow\uparrow\rangle)(|\uparrow\downarrow\rangle+|\downarrow\uparrow\rangle$) is used; in this case, the time evolving of the process is strictly confined in the $SU(2)$ symmetric sectors and data from the two methods are agree with each other well. Fig.\ref{fig1} just proves the correctness of the proposed ancilla TEBD algorithm.

In this paper, we will use ED to achieve the long time scale evolution for short chains and ancilla TEBD for short time evolution of large size models. 

\section{Mutifractality}\label{multi}
Dynamical behaviors of one system depend directly on the statistical properties of its single eigenstate; therefore, before the study of the entropy dynamics $S(t)$, we first discuss the multifractal properties of eigenstates in the bond disordered Heisenberg chain. The key proxy of multifractal properties is the multifractal dimension $D_q$ which can be extracted from the relation between the participation entropy
\begin{equation}\label{sq}
  S_q = \frac{1}{1-q}\ln\sum_{\alpha=1}^{D_H}|c_\alpha|^{2q}
\end{equation} 
and the size of the Hilbert space as mentioned in Sec.\ref{intro}; here, $|c_\alpha|$ denotes the expanding coefficient of the initial wave function in the configuration space. To pick up the proper target states, we put the energy window in the middle of the spectrum where the corresponding eigenvectors prefer to be delocalized; the localization of these states will guarantee the localization of the whole spectrum. 

\begin{figure}
  \centering
  \includegraphics[width=1.0\linewidth]{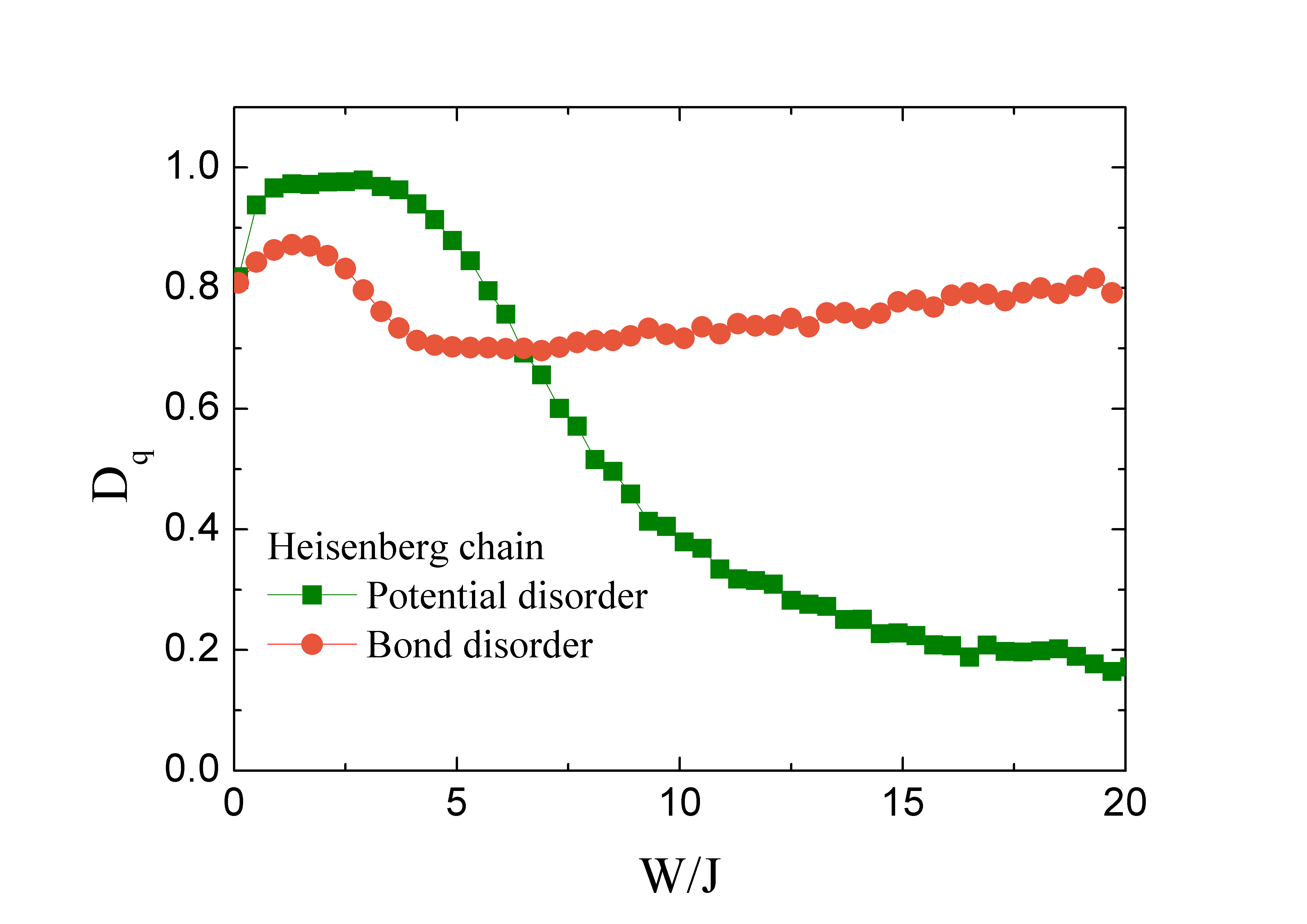}
  \caption{The dependence of the multifractal dimension $D_q$ on the disorder strength $W$. To make the comparison, we also plot the corresponding results of the potential disordered Heisenberg chain.}\label{Dq}
\end{figure}

We use ED method to obtain $D_q$ in the subspace with zero spin moments $S^{tot}_z=0$. First, we calculate the average of $S_q$ in the middle spectrum of spin chains with different lengths $L=8$, $10$, $12$, $14$; for each chain with fixed $L$, we use $500$ disorder samples to get the average of $S_q$. Then, we can extract $D_q$ from the slopping rate of $S_q\sim D_q\ln D_H$. We find that, as shown in Fig.\ref{Dq}, with the increasing of the disorder strength $W$, the multifractal dimension $D_q$ of bond disordered Heisenberg chain doesn't show the universal critical dropping as in the XXZ chain \cite{laflo19}. Since $D_q$ oscillate near $1$ and never drops from $1$ to $0$ as also plotted in Fig.\ref{Dq}, we may conclude that the eigenstates in the middle spectrum never localized and there is no MBL phase transition in the spectrum. This result of $D_q$ is also in line with the intuitive view of the bond disordered spin chain. It has been proved that bond disordered potential can not fully localize free particles since the localization length of eigenstates with energy $E\sim0$ diverges logarithmically.\cite{zy16} In addition, the spin interaction term $\Delta S^z_iS^z_j$ prefers not to induce localization; therefore, in this viewpoint, MBL is also not expected in the bond disordered Heisenberg chain.

\section{Entropy dynamics}\label{entropy}
\begin{figure}
  \centering
  \includegraphics[width=1.0\linewidth]{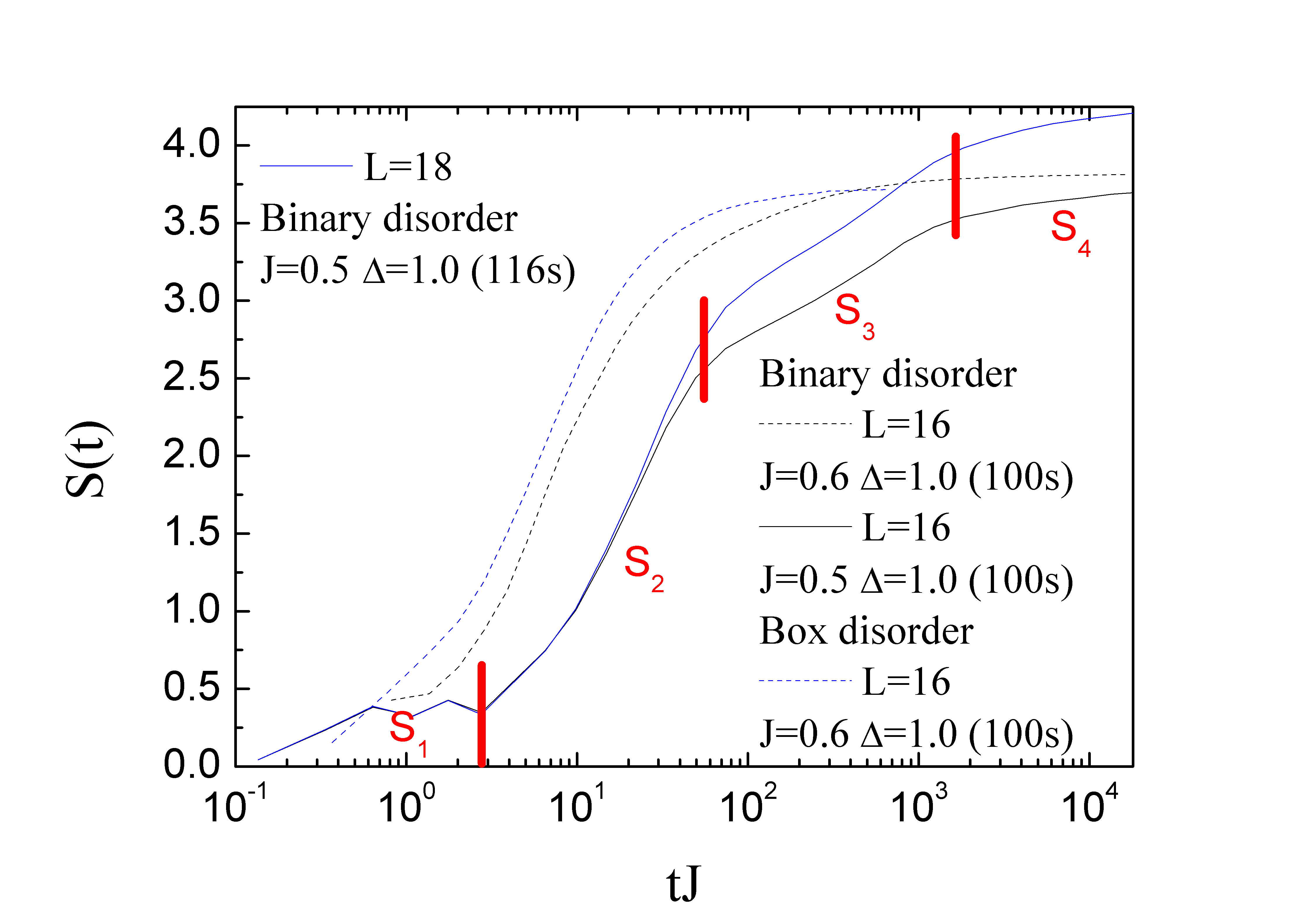}
  \caption{Entropy dynamics of bond disordered Heisenberg chains and XXZ chains with box and binary disorder. We set the disorder strength $W=1.8$ ; the number of samples is labeled in the legend.}\label{fig3}
\end{figure}

Although the multifractal dimension $D_q$ indicates the persistent of delocalization of the eigenstates in bond disordered Heisenberg model, interestingly, the von-Neumann entropy $S(t)$ can still exhibit a remarkable long time scaling behavior. As mentioned in Ref.[\onlinecite{zy16}], for the bond disordered Heisenberg model, $S(t)$ exhibit unusual long time scaling when $t>10^2$. Here, in Fig.\ref{fig3}, using ED method, we first reproduce the long time scaling behavior for larger chain length $L=16$ and $18$; 
We do the quantum quench starting from an initial Neel state. In contrast to the potential disordered spin chains, for the binary bond disordered cases, after the initial ballistic spreading and the universal oscillation ($0\le t\le t_1^\ast$), $S(t)$ goes up more faster till $t_2^\ast\sim 10^2$; this quick enhancement of the entanglement entropy make the long time evolution of the bond disordered chain intractable for TEBD simulation. Remarkably, $S(t)$ exhibits another slow long time scaling at $t>t_2^\ast$ until saturates at about $t=t_3^\ast$. Therefore, the entanglement dynamics of binary bond disordered Heisenberg model can be divided into four sectors: $S1$, $S2$, $S3$ and $S4$ which are divided sequentially by the three time scales $t_1^\ast$, $t_2^\ast$ and $t_3^\ast$. Next, we will concentrate on $S3$. 

Interestingly, Fig.\ref{fig3} also indicates that sector $S3$ only appears for the binary bonded disordered Heisenberg model. Setting $J=0.6$ and $L=16$, we break the $SU(2)$ symmetry and plot the corresponding $S(t)$ in the black dashed line. On the other hand, we also replace the binary disorder with the uniform box disorder and plot the result in the blue dashed line. Clearly, $S3$ disappears in these cases and indicates that $S3$ is adherent to the non-abelian symmetry. In order to detect the physical origin of the occurrence of $S3$, in the following, we calculate the time evolution of the initial state fidelity and inverse participation ratio to see what happens with the emergence of $S3$.   
\begin{figure}
  \centering
  \includegraphics[width=1.0\linewidth]{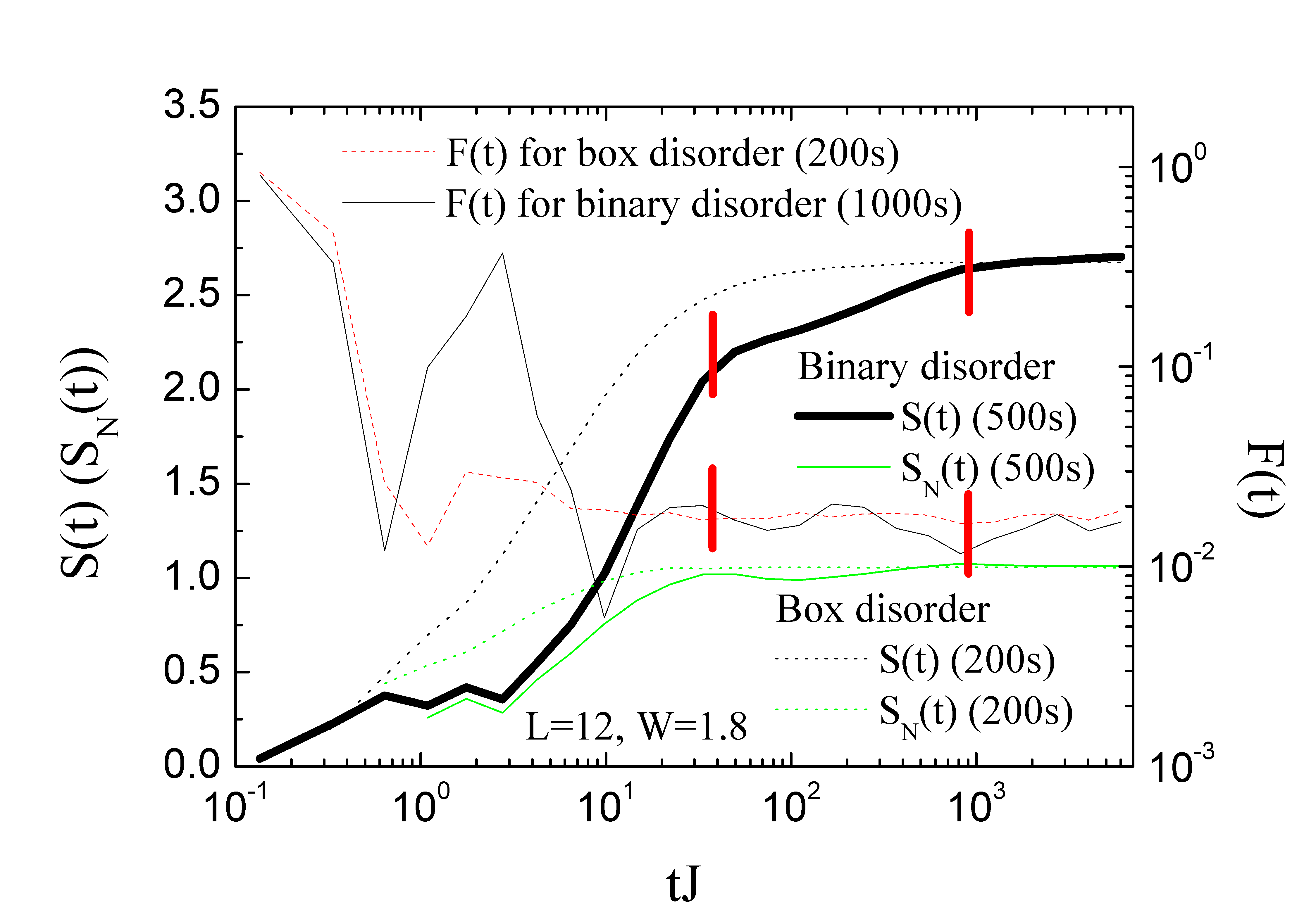}
  \caption{Fidelity of bond disordered Heisenberg chains with box and binary disorder $W=1.8$, $L=12$. $S(t)$ and number entropy $S_N(t)$ are also plotted to guide eyes.}\label{fig4}
\end{figure}

First, we study the dynamics of fidelity $F(t)=|\langle\Phi(0)|\Phi(t)\rangle|^2$. $F(t)$ describes the returning probability to the initial state; it usually decays to the averagement of IPR of the initial state $|\Phi(0)\rangle$\cite{Santos15}. Fig.\ref{fig4} shows that, with the occurrence of $S3$, $F(t)$ exhibits persistent oscillations which is absent for the box disorder potential; the oscillations along with $S3$ can not be averaged out by increasing the disorder samples. It is believed that the oscillations of $F(t)$ is in tight relation with the onset of LIOMs. Therefore, we may infer that $|\Phi(t)\rangle$ evolves to a transient state that produces the unusual long time scaling entropy dynamics and the dynamics is controlled by a set of LIMOs. To supplement the entropy spreading behavior, we also calculate the number entropy\cite{SN} $S_N=-\sum_np(n)\ln p(n)$, $p(n)$ is the probability to find $n$ particles or spins in the considered subsystem; the number entropy measure the particle fluctuations between subsystem and the environment. We can see in Fig.\ref{fig4} that $S_N$ saturates finally for both of the bond and box disordered Heisenberg chain. 
\begin{figure}
  \centering
  \includegraphics[width=1.0\linewidth]{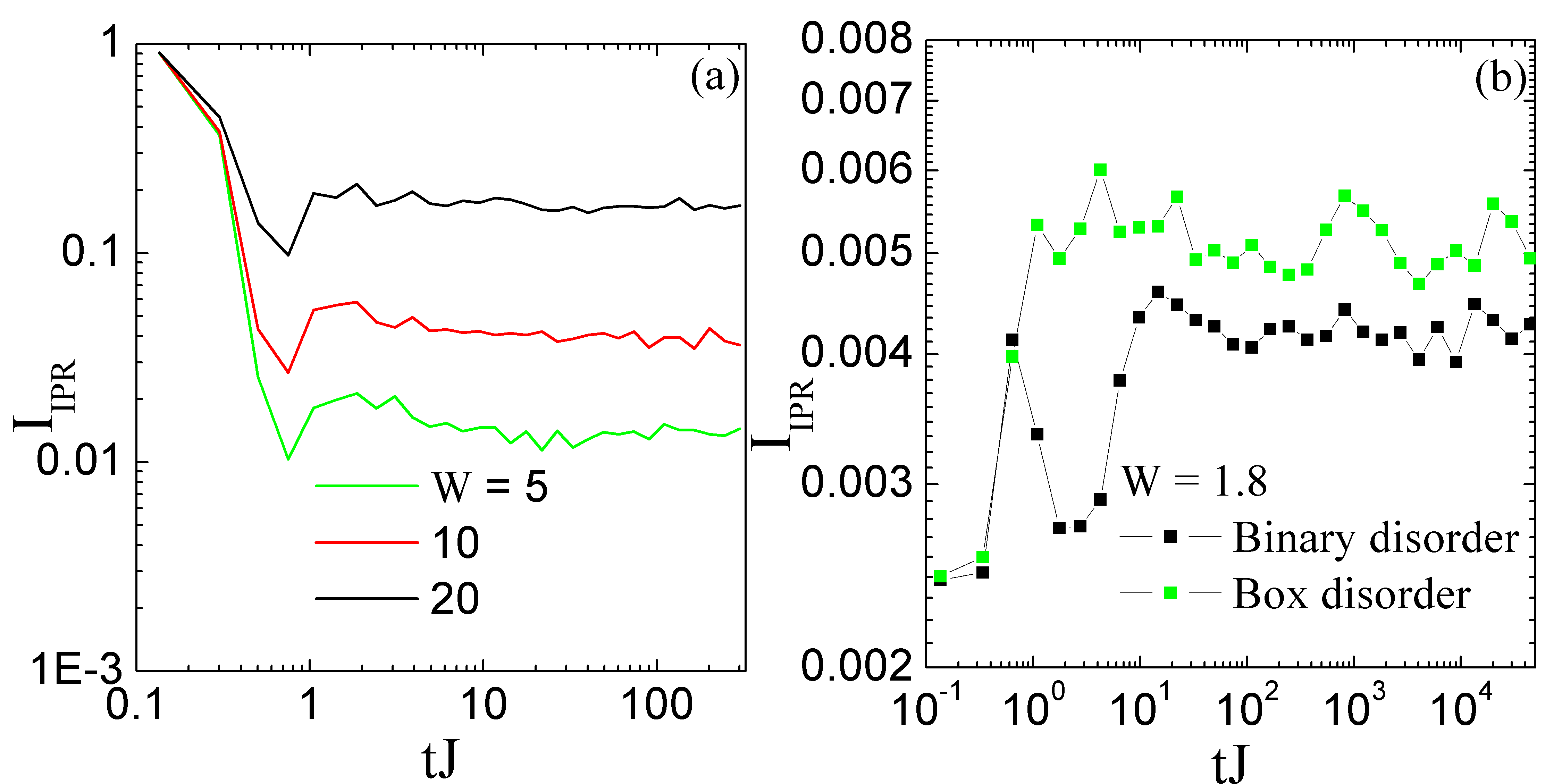}
  \caption{(a) shows the dynamics of the IPR $I_{IPR}(t)$ of potential disordered Heisenberg chain in the $U(1)$ basis with different box disorder $W$, $L=12$. (b) shows the dynamics of the IPR $I_{IPR}(t)$ of bond disordered Heisenberg chain in the $SU(2)$ basis with $W=1.8$, $L=12$. We use $100$ disorder samples to obtain each curve.}\label{fig5}
\end{figure}

To justify the localization of $|\Phi(t)\rangle$, we calculate its inverse participation ratio $I_{IPR}(t)$ at each time step. IPR describes the extendability of a quantum state in the Hilbert space with $D_H$ eigenbasis $|\psi_n\rangle$. Being expanded as $|\Phi(0)\rangle=\sum_{n}c_n|\psi_n\rangle$, then, the IPR can be defined as $I_{IPR}=\sum_{n}|c_n|^4$. It is easy to see that $I_{IPR}=1$ means $|\Phi(0)\rangle$ is localized and $I_{IPR}=1/D_H$ denotes a fully extended state. However, $I_{IPR}$ depends on the basis vectors of the Hilbert space; in this paper, we consider only two kinds of basis: the $U(1)$ basis consisting of the product states of local spin vectors $|\uparrow\rangle$ ($|\downarrow\rangle$) and the $SU(2)$ basis which is the linear combination of the $U(1)$ basis by using Clebsch-Gordan coefficients.  Fig.\ref{fig5} (a) shows that, for the potential disordered Heisenberg chain which obeys $U(1)$ symmetry, the converged value of $I_{IPR}(t)$ increases with the disorder strength $W$ indicating that the final state of the time evolution tends to be localized in the $U(1)$ subspace with $S_z^{tot}=\sum_jS_j^z=0$. For the bond disordered Heisenberg chain which obeys the $SU(2)$ symmetry, as shown in Fig.\ref{fig5} (b), the initial Neel state is quite an extended state in the $SU(2)$ basis with zero total spin magnetic moment that the initial $I_{IPR}(0)$ approaches to $1/924\simeq0.0011$ which is the lower limit of IPR for $L=12$. Then, during the time evolving process, $I_{IPR}(t)$ increases and keeps long time oscillations; In the long time limit, $I_{IPR}(t)$ generally doesn't exceed $10^{-2}$ which is very near to the lower limit. Consequently, we can conclude that the final state of long time evolution will not localize in the bond disordered Heisenberg model. 
\begin{figure}
  \centering
  \includegraphics[width=1.0\linewidth]{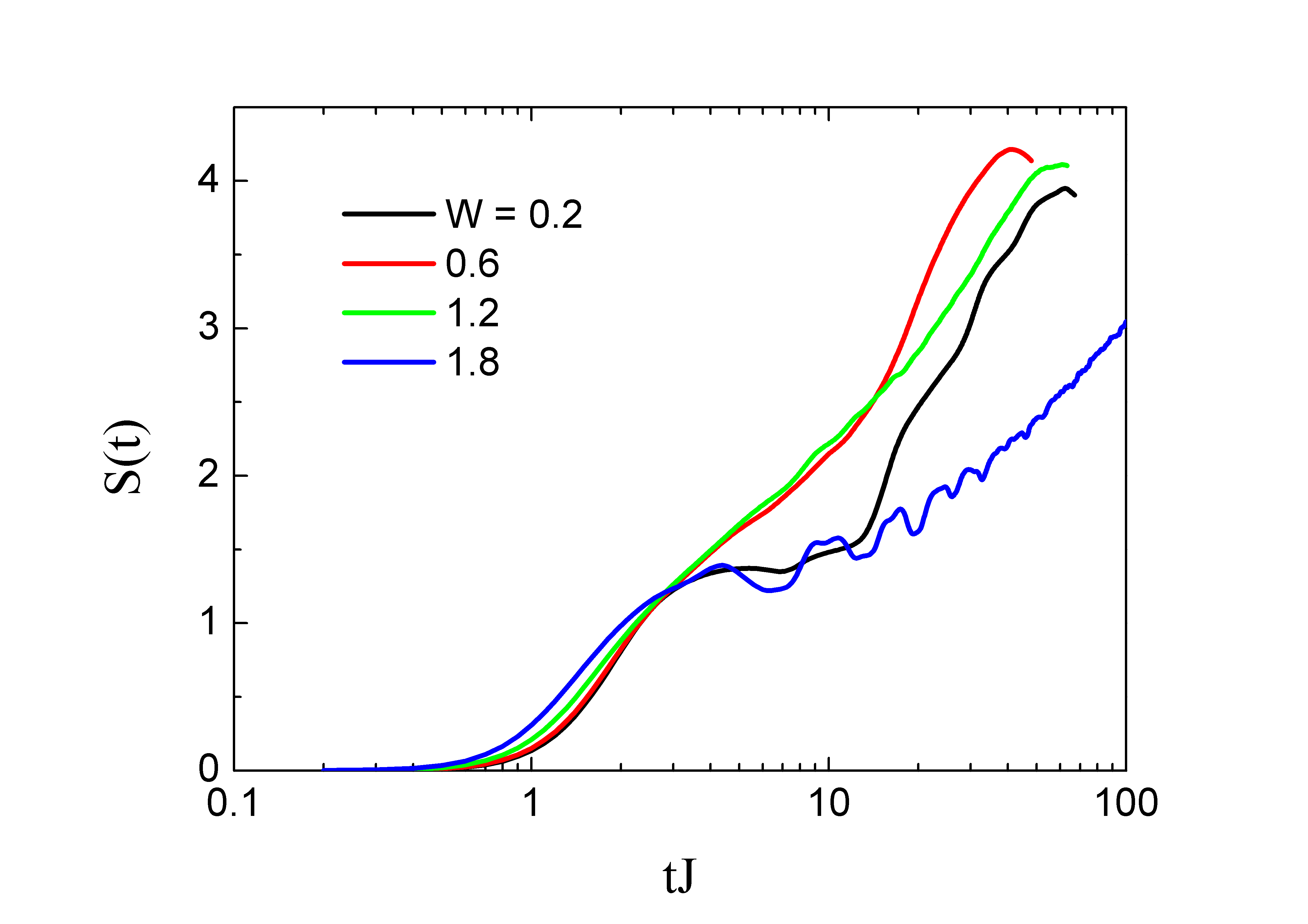}
  \caption{Ancilla TEBD results of entropy dynamics of bond disordered Heisenberg chains with different binary disorder $W$ and $L=30$ in the two-spin up subspace.}\label{fig6}
\end{figure} 
\begin{figure}
  \centering
  \includegraphics[width=1.0\linewidth]{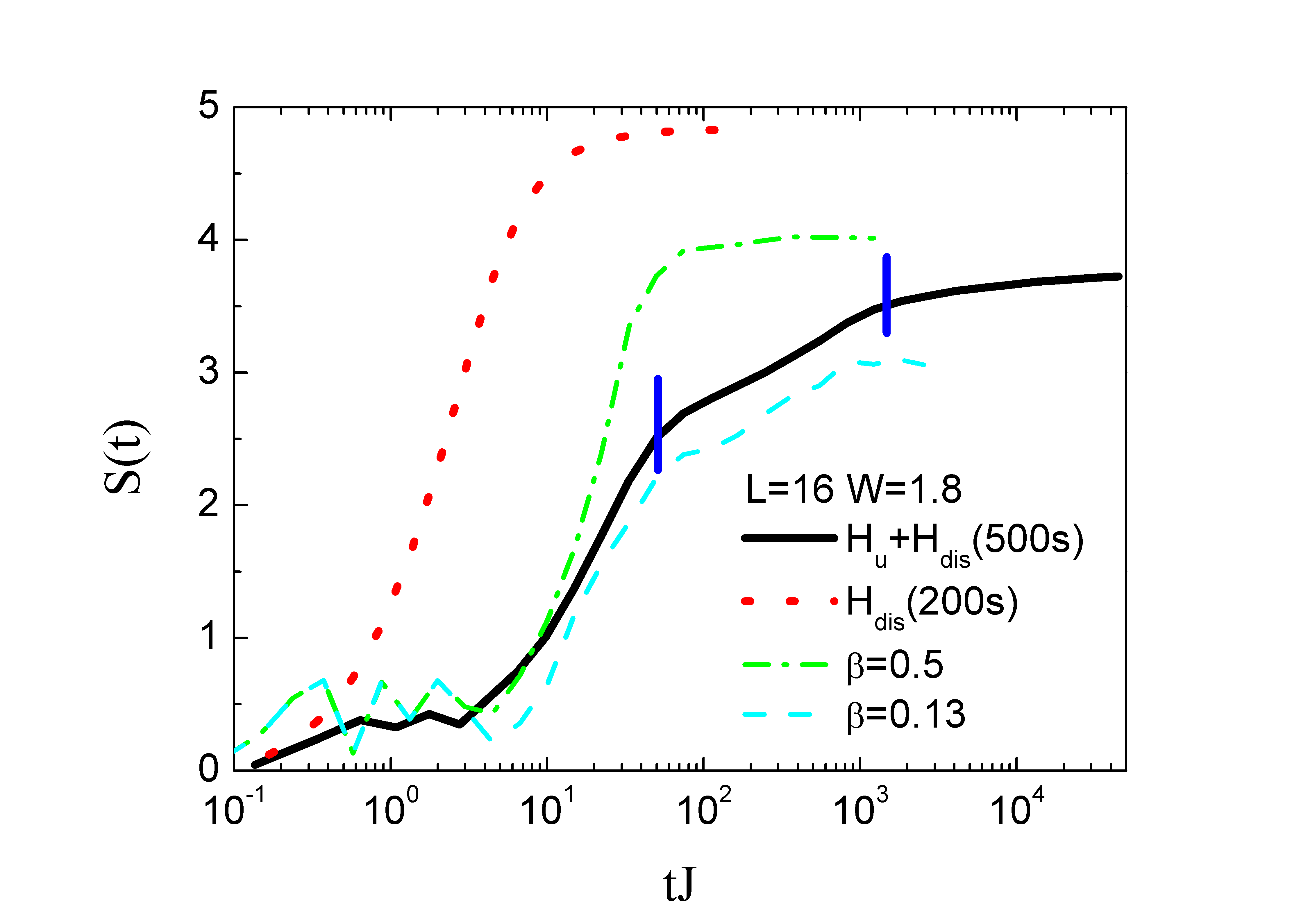}
  \caption{Entropy dynamics of different parts in the Hamiltonian of the bond disordered Heisenberg chain with binary disorder $W=1.8$ and the contributions from disorder samples with different $\beta$ are also shown.}\label{fig7}
\end{figure}

Since the final state of the quench in bond disordered Heisenberg chain is not a MBL state, to address the long time scaling of $S(t)$, we turn to study the competing effect of the spin interaction and the disorder strength $W$. We note that Eq.\ref{eq2} consists of two parts: the uniform part $H_u$ and the disordered part $H_{dis}$. Then, Eq.\ref{eq2} can be rewritten as
\begin{eqnarray}\label{eq4}
   H&=&H_u+H_{dis}\\
   H_u&=& \sum_i[J(S^{+}_iS^{-}_{i+1} + h.c.)+\Delta S^z_iS^z_{i+1}]\nonumber\\
   H_{dis}&=&\sum_i\frac{W}{2}[J(\sigma^z_iS^{+}_iS^{-}_{i+1} + h.c.)+\Delta \sigma^z_iS^z_iS^z_{i+1}].\nonumber
\end{eqnarray} 
$H_u$ and $H_{dis}$ contribute to $S(t)$ in different ways. To supplement the ED results and consider more disorder samples, by using the ancilla TEBD, we calculated the entropy dynamics in the subspace with $S^{tot}_z=2-L/2$ in which only two up spins exist so that we can study the dynamical behavior of interacting models with longer chain length $L=30$; the disorder sample number is $2^{29}$. Here, we calculate the von Neumann entropy $S(t)$ by considering half of the sites in the MPS as the subsystem including the ancilla sites; it should be stressed that $S(t)$ defined in this manner is actually not the actual von Neumann entropy of the subsystem but $S(t)$ will show the same scaling behavior and is easy to be calculated by grouping the singular values at the cutting point of the MPS. Fig.\ref{fig6} shows that $S(t)$ is dominated by $H_u$ at short time $t\le 1/J$ when $W<J$ and $S(t)$ exhibits the initial linear increasing behavior; later, at about $t\sim J/W^2$, the disorder part $H_{dis}$ takes part in and produces a logarithmic like scaling behavior. $S(t)$ saturates quickly since the whole system is dominated by $H_u$. With the increasing of $H_{dis}$, the disorder part becomes more significant and produces a well behaved logarithmic scaling at $t> 10^2$ when $W=1.8$; But, when $t\sim10$, there exists a transient step that separates the initial linear scaling and later logarithmic like scaling. This kind of separation of the dynamical process seems quite similar to the entropy dynamics in MBL systems; however, it s apparently not true for the bond disordered Heisenberg model since the existence of MBL states has been ruled out. 

In order to explain the unusual entropy dynamic in Fig.\ref{fig6}, using ED, we study the contributions of different binary disorder samples as shown in Eq.\ref{eq4}. Due to the properties of the binary disorder, we see that the disorder potential consist of randomly distributed strong bonds with strength $J(1+ W/2)$($\Delta(1+ W/2)$) and weak bond with $J(1- W/2)$ ($\Delta(1- W/2)$) in which $W/2<1$; we choose $W=1.8$ here. Define the ratio of the number of strong bonds to the weak bonds as $\beta$; we can see clearly in Fig.\ref{fig7} that samples with different $\beta$ contributes differently to the entropy dynamics. Initially, the dynamical behavior of $S(t)$ is dominated by the strong bonds. $S(t)$ show a quick initial increasing and oscillates at about $t\sim10^0$. Later, the dynamics of $S(t)$ is controlled by the weak bonds and the time scale is set by $1/[J(1- W/2)]$. But $S(t)$ behaves differently for different $\beta$. As shown in Fig.\ref{fig7}, for $\beta=0.5$, $S(t)$ goes up quickly and saturates before $t\sim10^2$; while for $\beta=0.13$, due to the large portion of weak bonds, $S(t)$ keeps the increasing behavior till $t\sim10^3$ during which the line shape of $S(t)$ depends sensitively on the distribution of the strong bonds around the cutting position of the subsystem and the environment; here, we may also expect that the acting of LIMOs contained in the weak bonds contributes to the persistent oscillation of $F(t)$ at the same time scale. Consequently, we can see that all kinds of disorder samples make a contribution to $S(t)$ when $t<10^2$ and only samples with small $\beta$ contribute to the later time dynamics when $t>10^2$. After doing the disorder average, we obtain finally the expected dynamical scaling behavior with the aforementioned four sectors.  

To give a further justification, we also plot the entropy dynamics governed by only the disordered part $H_{dis}$ in Eq.\ref{eq4} in Fig.\ref{fig7} as the red dashed line. Clearly, $S(t)$ saturates at about $t\sim10$ and does not show the four dynamical sectors. At last, we can see that the special long time scaling behavior of the entropy dynamics of the bond disordered Heisenberg model is due to the competing effects of the spin interaction and the strength of the disorder potential which give a new mechanism for producing long time entropy dynamics. 

It should be noted that, for the uniform box disorder, due to continuous distribution of $\beta$ among the infinite number of disorder samples, the differences between the entropy dynamics associated to different $\beta$ are smeared out; In consequence, one may not see the four-sector dynamical configuration of the entropy dynamics in this scenario.  

\begin{figure}
  \centering
  \includegraphics[width=1.0\linewidth]{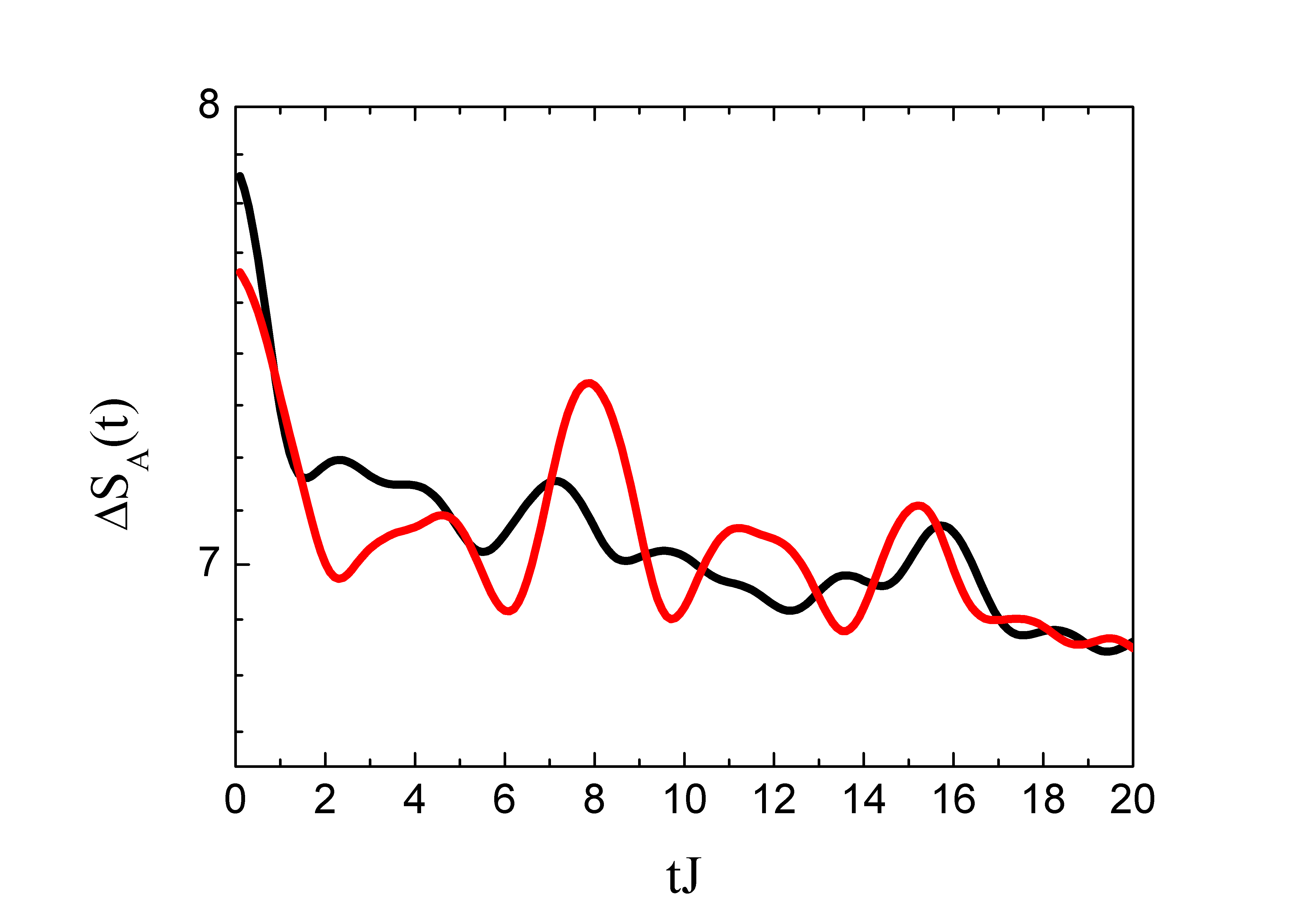}
  \caption{Entanglement asymmetry of bond disordered Heisenberg chains with binary disorder $W=1.8$, $L=100$.}\label{fig8}
\end{figure}
\section{Transient Mpemba effect}\label{Mpemba}
At last, considering the existence of many non-ergodic eigenstates in a non-abelian symmetric model and the Mpemba effect in spin glass models\cite{Mpembag}, we also study the decaying behavior of an initially nonsymmetric state. Technically, using the ancilla TEBD method, we randomly generate two initial MPS which are far from their corresponding long time steady symmetrical states. To measure the extent of symmetry breaking of the initial state, we adopt the entanglement asymmetry\cite{Mpemba1} $\Delta S_A=Tr(\rho_s\ln\rho_s-\rho^Q_s\ln\rho^Q_s)$, where $\rho_s$ is the reduced density matrix of the subsystem; $\rho^Q_s$ is the reduced density matrix in the projected symmetric subspaces with charge $Q$ such as the total spin moments. Intuitively, the quantum quench will drive the initial unsymmetric state to its stationary state that fulfills the symmetry of the Hamiltonian; states with small $\Delta S_A$ always decays faster. Although the ancilla TEBD can only give the short time dynamics of a finite system, the decaying of $\Delta S_A$ still give some hints for the existing of the quantum Mpemba effect. See Fig.\ref{fig8},  counter intuitively, in the bond disordered Heisenberg model, the initially more asymmetric state may approach to the final symmetric stationary state more faster at least in some time domains because of the inter crossings of the two curves. We name such phenomena as transient quantum Mpemba effect (TQME). Here, we perform the time evolution on a chain with length $L=100$ to avoid the finite size effect at the short time regime. It should be stressed that Fig.\ref{fig8} only supports the existence of two states in the system that will exhibit TQME; we are not mean every two initial states shows TQME. Further, our calculations does not rule out the existence of TQME in other kinds of disordered models.   

\section{Conclusion}\label{conclude}
In this article, we give a thorough study of the dynamical properties of the bond disordered Heisenberg model. At first, we propose an efficient algorithm, the ancilla TEBD method which combines the purification scheme with the TEBD algorithms, to simulate the entropy dynamics of bond disordered model. With the support of ED calculation, we obtain the multifractal dimension $D_q$ of the bond disordered Heisenberg chain and find no critical behavior with the increasing of disorder strength which agrees with the absence of MBL in non-ableian systems. To explain the occurrence of the long time logarithmic scaling of $S(t)$ at $t\sim10^2$, based on our ED and ancilla TEBD calculation, we confirm that the long time scaling is induced by the competition of the spin interaction and the disorder. Our numerical results propose a new physical mechanism other than MBL to produce long time slow scaling behavior of the entanglement quench dynamics. Such a long time scaling may be related to the symmetrical properties of both of the uniform part $H_u$ and disordered part $H_{dis}$. At last, we also study the quantum Mpemba effect in the bond disordered spin chains by calculating the entanglement asymmetry $\Delta S_A$ and our results support the existence of the transient quantum Mpemba effect, but the destiny of the decaying of $\Delta S_A$ needs further study.   
\section*{acknowledgments}
We thank Prof. Jesko Sirker for very useful discussions. Yang Zhao acknowledges support from the Natural Science Basic Research Plan in Shaanxi Province of China under Grant No. 2021JM-041 and Hebei Normal University under Grant No. L2023B06.
\bibliography{ancibib}

\begin{thebibliography}{20}
\expandafter\ifx\csname natexlab\endcsname\relax\def\natexlab#1{#1}\fi
\expandafter\ifx\csname bibnamefont\endcsname\relax
  \def\bibnamefont#1{#1}\fi
\expandafter\ifx\csname bibfnamefont\endcsname\relax
  \def\bibfnamefont#1{#1}\fi
\expandafter\ifx\csname citenamefont\endcsname\relax
  \def\citenamefont#1{#1}\fi
\expandafter\ifx\csname url\endcsname\relax
  \def\url#1{\texttt{#1}}\fi
\expandafter\ifx\csname urlprefix\endcsname\relax\def\urlprefix{URL }\fi
\providecommand{\bibinfo}[2]{#2}
\providecommand{\eprint}[2][]{\url{#2}}

\bibitem[{\citenamefont{Doggen et~al.}(2020)\citenamefont{Doggen, Gornyi,
  Mirlin, and Polyakov}}]{PImb}
\bibinfo{author}{\bibfnamefont{E.~V.~H.} \bibnamefont{Doggen}},
  \bibinfo{author}{\bibfnamefont{I.~V.} \bibnamefont{Gornyi}},
  \bibinfo{author}{\bibfnamefont{A.~D.} \bibnamefont{Mirlin}},
  \bibnamefont{and} \bibinfo{author}{\bibfnamefont{D.~G.}
  \bibnamefont{Polyakov}}, \bibinfo{journal}{Phys. Rev. Lett.}
  \textbf{\bibinfo{volume}{125}}, \bibinfo{pages}{155701}
  (\bibinfo{year}{2020}),
  \urlprefix\url{https://link.aps.org/doi/10.1103/PhysRevLett.125.155701}.

\bibitem[{\citenamefont{Potter and Vasseur}(2016)}]{0su2}
\bibinfo{author}{\bibfnamefont{A.~C.} \bibnamefont{Potter}} \bibnamefont{and}
  \bibinfo{author}{\bibfnamefont{R.}~\bibnamefont{Vasseur}},
  \bibinfo{journal}{Phys. Rev. B} \textbf{\bibinfo{volume}{94}},
  \bibinfo{pages}{224206} (\bibinfo{year}{2016}),
  \urlprefix\url{https://link.aps.org/doi/10.1103/PhysRevB.94.224206}.

\bibitem[{\citenamefont{Protopopov et~al.}(2017)\citenamefont{Protopopov, Ho,
  and Abanin}}]{1stsu2}
\bibinfo{author}{\bibfnamefont{I.~V.} \bibnamefont{Protopopov}},
  \bibinfo{author}{\bibfnamefont{W.~W.} \bibnamefont{Ho}}, \bibnamefont{and}
  \bibinfo{author}{\bibfnamefont{D.~A.} \bibnamefont{Abanin}},
  \bibinfo{journal}{Phys. Rev. B} \textbf{\bibinfo{volume}{96}},
  \bibinfo{pages}{041122} (\bibinfo{year}{2017}),
  \urlprefix\url{https://link.aps.org/doi/10.1103/PhysRevB.96.041122}.

\bibitem[{\citenamefont{Protopopov et~al.}(2020)\citenamefont{Protopopov,
  Panda, Parolini, Scardicchio, Demler, and Abanin}}]{Abaninprx}
\bibinfo{author}{\bibfnamefont{I.~V.} \bibnamefont{Protopopov}},
  \bibinfo{author}{\bibfnamefont{R.~K.} \bibnamefont{Panda}},
  \bibinfo{author}{\bibfnamefont{T.}~\bibnamefont{Parolini}},
  \bibinfo{author}{\bibfnamefont{A.}~\bibnamefont{Scardicchio}},
  \bibinfo{author}{\bibfnamefont{E.}~\bibnamefont{Demler}}, \bibnamefont{and}
  \bibinfo{author}{\bibfnamefont{D.~A.} \bibnamefont{Abanin}},
  \bibinfo{journal}{Phys. Rev. X} \textbf{\bibinfo{volume}{10}},
  \bibinfo{pages}{011025} (\bibinfo{year}{2020}),
  \urlprefix\url{https://link.aps.org/doi/10.1103/PhysRevX.10.011025}.

\bibitem[{\citenamefont{Saraidaris et~al.}(2024)\citenamefont{Saraidaris, Li,
  Weichselbaum, von Delft, and Abanin}}]{andreas2024}
\bibinfo{author}{\bibfnamefont{D.}~\bibnamefont{Saraidaris}},
  \bibinfo{author}{\bibfnamefont{J.-W.} \bibnamefont{Li}},
  \bibinfo{author}{\bibfnamefont{A.}~\bibnamefont{Weichselbaum}},
  \bibinfo{author}{\bibfnamefont{J.}~\bibnamefont{von Delft}},
  \bibnamefont{and} \bibinfo{author}{\bibfnamefont{D.~A.}
  \bibnamefont{Abanin}}, \bibinfo{journal}{Phys. Rev. B}
  \textbf{\bibinfo{volume}{109}}, \bibinfo{pages}{094201}
  (\bibinfo{year}{2024}),
  \urlprefix\url{https://link.aps.org/doi/10.1103/PhysRevB.109.094201}.

\bibitem[{\citenamefont{Vasseur et~al.}(2015)\citenamefont{Vasseur, Potter, and
  Parameswaran}}]{qcg}
\bibinfo{author}{\bibfnamefont{R.}~\bibnamefont{Vasseur}},
  \bibinfo{author}{\bibfnamefont{A.~C.} \bibnamefont{Potter}},
  \bibnamefont{and} \bibinfo{author}{\bibfnamefont{S.~A.}
  \bibnamefont{Parameswaran}}, \bibinfo{journal}{Phys. Rev. Lett.}
  \textbf{\bibinfo{volume}{114}}, \bibinfo{pages}{217201}
  (\bibinfo{year}{2015}),
  \urlprefix\url{https://link.aps.org/doi/10.1103/PhysRevLett.114.217201}.

\bibitem[{\citenamefont{Aramthottil et~al.}(2024)\citenamefont{Aramthottil,
  Sierant, Lewenstein, and Zakrzewski}}]{XXZMBL2024}
\bibinfo{author}{\bibfnamefont{A.~S.} \bibnamefont{Aramthottil}},
  \bibinfo{author}{\bibfnamefont{P.}~\bibnamefont{Sierant}},
  \bibinfo{author}{\bibfnamefont{M.}~\bibnamefont{Lewenstein}},
  \bibnamefont{and}
  \bibinfo{author}{\bibfnamefont{J.}~\bibnamefont{Zakrzewski}},
  \bibinfo{journal}{Phys. Rev. Lett.} \textbf{\bibinfo{volume}{133}},
  \bibinfo{pages}{196302} (\bibinfo{year}{2024}),
  \urlprefix\url{https://link.aps.org/doi/10.1103/PhysRevLett.133.196302}.

\bibitem[{\citenamefont{Dey et~al.}(2020)\citenamefont{Dey, Andrade, and
  Vojta}}]{Vojtaprb}
\bibinfo{author}{\bibfnamefont{S.}~\bibnamefont{Dey}},
  \bibinfo{author}{\bibfnamefont{E.~C.} \bibnamefont{Andrade}},
  \bibnamefont{and} \bibinfo{author}{\bibfnamefont{M.}~\bibnamefont{Vojta}},
  \bibinfo{journal}{Phys. Rev. B} \textbf{\bibinfo{volume}{102}},
  \bibinfo{pages}{125121} (\bibinfo{year}{2020}),
  \urlprefix\url{https://link.aps.org/doi/10.1103/PhysRevB.102.125121}.

\bibitem[{\citenamefont{Zhao et~al.}(2016)\citenamefont{Zhao, Andraschko, and
  Sirker}}]{zy16}
\bibinfo{author}{\bibfnamefont{Y.}~\bibnamefont{Zhao}},
  \bibinfo{author}{\bibfnamefont{F.}~\bibnamefont{Andraschko}},
  \bibnamefont{and} \bibinfo{author}{\bibfnamefont{J.}~\bibnamefont{Sirker}},
  \bibinfo{journal}{Phys. Rev. B} \textbf{\bibinfo{volume}{93}},
  \bibinfo{pages}{205146} (\bibinfo{year}{2016}),
  \urlprefix\url{https://link.aps.org/doi/10.1103/PhysRevB.93.205146}.

\bibitem[{\citenamefont{Andraschko et~al.}(2014)\citenamefont{Andraschko, Enss,
  and Sirker}}]{felix}
\bibinfo{author}{\bibfnamefont{F.}~\bibnamefont{Andraschko}},
  \bibinfo{author}{\bibfnamefont{T.}~\bibnamefont{Enss}}, \bibnamefont{and}
  \bibinfo{author}{\bibfnamefont{J.}~\bibnamefont{Sirker}},
  \bibinfo{journal}{Phys. Rev. Lett.} \textbf{\bibinfo{volume}{113}},
  \bibinfo{pages}{217201} (\bibinfo{year}{2014}),
  \urlprefix\url{https://link.aps.org/doi/10.1103/PhysRevLett.113.217201}.

\bibitem[{\citenamefont{Rylands et~al.}(2024)\citenamefont{Rylands, Klobas,
  Ares, Calabrese, Murciano, and Bertini}}]{Mpemba1}
\bibinfo{author}{\bibfnamefont{C.}~\bibnamefont{Rylands}},
  \bibinfo{author}{\bibfnamefont{K.}~\bibnamefont{Klobas}},
  \bibinfo{author}{\bibfnamefont{F.}~\bibnamefont{Ares}},
  \bibinfo{author}{\bibfnamefont{P.}~\bibnamefont{Calabrese}},
  \bibinfo{author}{\bibfnamefont{S.}~\bibnamefont{Murciano}}, \bibnamefont{and}
  \bibinfo{author}{\bibfnamefont{B.}~\bibnamefont{Bertini}},
  \bibinfo{journal}{Phys. Rev. Lett.} \textbf{\bibinfo{volume}{133}},
  \bibinfo{pages}{010401} (\bibinfo{year}{2024}),
  \urlprefix\url{https://link.aps.org/doi/10.1103/PhysRevLett.133.010401}.

\bibitem[{\citenamefont{Baity-Jesi et~al.}(2019)\citenamefont{Baity-Jesi,
  Calore, Cruz, Fernandez, Gil-Narvion, Gordillo-Guerrero, Iniguez, Lasanta,
  Maiorano, Marinari et~al.}}]{Mpembag}
\bibinfo{author}{\bibfnamefont{M.}~\bibnamefont{Baity-Jesi}},
  \bibinfo{author}{\bibfnamefont{E.}~\bibnamefont{Calore}},
  \bibinfo{author}{\bibfnamefont{A.}~\bibnamefont{Cruz}},
  \bibinfo{author}{\bibfnamefont{L.~A.} \bibnamefont{Fernandez}},
  \bibinfo{author}{\bibfnamefont{J.~M.} \bibnamefont{Gil-Narvion}},
  \bibinfo{author}{\bibfnamefont{A.}~\bibnamefont{Gordillo-Guerrero}},
  \bibinfo{author}{\bibfnamefont{D.}~\bibnamefont{Iniguez}},
  \bibinfo{author}{\bibfnamefont{A.}~\bibnamefont{Lasanta}},
  \bibinfo{author}{\bibfnamefont{A.}~\bibnamefont{Maiorano}},
  \bibinfo{author}{\bibfnamefont{E.}~\bibnamefont{Marinari}},
  \bibnamefont{et~al.}, \textbf{\bibinfo{volume}{116}}, \bibinfo{pages}{15350}
  (\bibinfo{year}{2019}),
  \eprint{https://www.pnas.org/doi/pdf/10.1073/pnas.1819803116},
  \urlprefix\url{https://www.pnas.org/doi/abs/10.1073/pnas.1819803116}.

\bibitem[{\citenamefont{De~Tomasi and Khaymovich}(2020)}]{MuF3}
\bibinfo{author}{\bibfnamefont{G.}~\bibnamefont{De~Tomasi}} \bibnamefont{and}
  \bibinfo{author}{\bibfnamefont{I.~M.} \bibnamefont{Khaymovich}},
  \bibinfo{journal}{Phys. Rev. Lett.} \textbf{\bibinfo{volume}{124}},
  \bibinfo{pages}{200602} (\bibinfo{year}{2020}),
  \urlprefix\url{https://link.aps.org/doi/10.1103/PhysRevLett.124.200602}.

\bibitem[{\citenamefont{Duthie et~al.}(2022)\citenamefont{Duthie, Roy, and
  Logan}}]{MuF2}
\bibinfo{author}{\bibfnamefont{A.}~\bibnamefont{Duthie}},
  \bibinfo{author}{\bibfnamefont{S.}~\bibnamefont{Roy}}, \bibnamefont{and}
  \bibinfo{author}{\bibfnamefont{D.~E.} \bibnamefont{Logan}},
  \bibinfo{journal}{Phys. Rev. B} \textbf{\bibinfo{volume}{106}},
  \bibinfo{pages}{L020201} (\bibinfo{year}{2022}),
  \urlprefix\url{https://link.aps.org/doi/10.1103/PhysRevB.106.L020201}.

\bibitem[{\citenamefont{Luitz et~al.}(2020)\citenamefont{Luitz, Khaymovich, and
  Lev}}]{MuF4}
\bibinfo{author}{\bibfnamefont{D.~J.} \bibnamefont{Luitz}},
  \bibinfo{author}{\bibfnamefont{I.~M.} \bibnamefont{Khaymovich}},
  \bibnamefont{and} \bibinfo{author}{\bibfnamefont{Y.~B.} \bibnamefont{Lev}},
  \bibinfo{journal}{SciPost Phys. Core} \textbf{\bibinfo{volume}{2}},
  \bibinfo{pages}{006} (\bibinfo{year}{2020}),
  \urlprefix\url{https://scipost.org/10.21468/SciPostPhysCore.2.2.006}.

\bibitem[{\citenamefont{Atas and Bogomolny}(2014)}]{MuF1}
\bibinfo{author}{\bibfnamefont{Y.~Y.} \bibnamefont{Atas}} \bibnamefont{and}
  \bibinfo{author}{\bibfnamefont{E.}~\bibnamefont{Bogomolny}},
  \bibinfo{journal}{Philosophical Transactions of the Royal Society A:
  Mathematical, Physical and Engineering Sciences}
  \textbf{\bibinfo{volume}{372}}, \bibinfo{pages}{20120520}
  (\bibinfo{year}{2014}),
  \eprint{https://royalsocietypublishing.org/doi/pdf/10.1098/rsta.2012.0520},
  \urlprefix\url{https://royalsocietypublishing.org/doi/abs/10.1098/rsta.2012.0520}.

\bibitem[{\citenamefont{Ketzmerick et~al.}(1997)\citenamefont{Ketzmerick,
  Kruse, Kraut, and Geisel}}]{Geisel97}
\bibinfo{author}{\bibfnamefont{R.}~\bibnamefont{Ketzmerick}},
  \bibinfo{author}{\bibfnamefont{K.}~\bibnamefont{Kruse}},
  \bibinfo{author}{\bibfnamefont{S.}~\bibnamefont{Kraut}}, \bibnamefont{and}
  \bibinfo{author}{\bibfnamefont{T.}~\bibnamefont{Geisel}},
  \bibinfo{journal}{Phys. Rev. Lett.} \textbf{\bibinfo{volume}{79}},
  \bibinfo{pages}{1959} (\bibinfo{year}{1997}),
  \urlprefix\url{https://link.aps.org/doi/10.1103/PhysRevLett.79.1959}.

\bibitem[{\citenamefont{Mac\'e et~al.}(2019)\citenamefont{Mac\'e, Alet, and
  Laflorencie}}]{laflo19}
\bibinfo{author}{\bibfnamefont{N.}~\bibnamefont{Mac\'e}},
  \bibinfo{author}{\bibfnamefont{F.}~\bibnamefont{Alet}}, \bibnamefont{and}
  \bibinfo{author}{\bibfnamefont{N.}~\bibnamefont{Laflorencie}},
  \bibinfo{journal}{Phys. Rev. Lett.} \textbf{\bibinfo{volume}{123}},
  \bibinfo{pages}{180601} (\bibinfo{year}{2019}),
  \urlprefix\url{https://link.aps.org/doi/10.1103/PhysRevLett.123.180601}.

\bibitem[{\citenamefont{Torres-Herrera and Santos}(2015)}]{Santos15}
\bibinfo{author}{\bibfnamefont{E.~J.} \bibnamefont{Torres-Herrera}}
  \bibnamefont{and} \bibinfo{author}{\bibfnamefont{L.~F.}
  \bibnamefont{Santos}}, \bibinfo{journal}{Phys. Rev. B}
  \textbf{\bibinfo{volume}{92}}, \bibinfo{pages}{014208}
  (\bibinfo{year}{2015}),
  \urlprefix\url{https://link.aps.org/doi/10.1103/PhysRevB.92.014208}.

\bibitem[{\citenamefont{Kiefer-Emmanouilidis
  et~al.}(2020)\citenamefont{Kiefer-Emmanouilidis, Unanyan, Fleischhauer, and
  Sirker}}]{SN}
\bibinfo{author}{\bibfnamefont{M.}~\bibnamefont{Kiefer-Emmanouilidis}},
  \bibinfo{author}{\bibfnamefont{R.}~\bibnamefont{Unanyan}},
  \bibinfo{author}{\bibfnamefont{M.}~\bibnamefont{Fleischhauer}},
  \bibnamefont{and} \bibinfo{author}{\bibfnamefont{J.}~\bibnamefont{Sirker}},
  \bibinfo{journal}{Phys. Rev. Lett.} \textbf{\bibinfo{volume}{124}},
  \bibinfo{pages}{243601} (\bibinfo{year}{2020}),
  \urlprefix\url{https://link.aps.org/doi/10.1103/PhysRevLett.124.243601}.

\end{thebibliography}
\end{document}